\documentclass[twocolumn,showpacs,amsmath,amssymb,nofootinbib,prd]{revtex4}
\usepackage{graphicx}
\usepackage{epsfig}
\usepackage{rotate}
\usepackage{amsmath}
\usepackage{amssymb}
\usepackage{amsfonts}
\usepackage{hyperref}
\usepackage{bm}

\usepackage[usenames]{color}
\usepackage{url}

\hypersetup{
    colorlinks=true,
    linkcolor=red,
    citecolor=blue,
}

\newcommand\ba{\begin{eqnarray}}
\newcommand\ea{\end{eqnarray}}
\newcommand\be{\begin{equation}}
\newcommand\ee{\end{equation}}
\renewcommand\H{{\cal H}}

\def\f{\frac}

\begin{document}

\title{Relativistic corrections and non-Gaussianity in radio continuum surveys}

\author{Roy Maartens$^{1,2}$, Gong-Bo Zhao$^{2,3}$, David Bacon$^2$,
Kazuya Koyama$^2$, Alvise Raccanelli$^{4,5}$\\~}

\affiliation{$^1$Physics Department, University of the Western
Cape, Cape Town 7535, South Africa\\ $^2$Institute of Cosmology \&
Gravitation,
University of Portsmouth, Portsmouth PO1 3FX, UK \\
$^3$National Astronomy Observatories, Chinese Academy of Science, Beijing, 100012, P.R.China\\
$^4$Jet Propulsion Laboratory, California Institute of Technology,
Pasadena CA 91109, US \\
$^5$California Institute of Technology, Pasadena CA 91125, US}

\date{\today}

\begin{abstract}

\noindent
Forthcoming radio continuum surveys will cover large volumes of the observable Universe and will reach to high redshifts, making them potentially powerful probes of dark
energy, modified gravity and non-Gaussianity. 
We
consider the continuum surveys with LOFAR, WSRT and ASKAP, and examples of continuum surveys with
the SKA. 
We extend recent work on these surveys by including redshift space distortions and lensing convergence in the radio source auto-correlation.
In addition we compute the general relativistic (GR) corrections to
the angular power spectrum. These GR corrections to the standard
Newtonian analysis of the power spectrum become significant on
scales near and beyond the Hubble scale at each redshift. 
We find that the GR corrections are at most percent-level in LOFAR, WODAN and EMU surveys, but they can produce
$O(10\%)$ changes for high enough sensitivity SKA continuum surveys. 
The signal is however dominated by cosmic variance, and multiple-tracer techniques will be needed to overcome this problem. 
The GR corrections are suppressed in continuum surveys because of the
integration over redshift -- we expect that GR corrections will be enhanced for future SKA HI surveys in which the source redshifts will be known.
We also provide predictions for the angular power spectra in the case where the primordial
perturbations have local non-Gaussianity. 
We find that non-Gaussianity dominates over GR corrections, and rises above cosmic variance when $f_{\rm NL}\gtrsim5$ for SKA continuum surveys.

\end{abstract}

\maketitle

\section{Introduction}

Radio continuum surveys for cosmology are entering a new phase,
given the imminent surveys with LOFAR (the LOw Frequency ARray for radio astronomy
\cite{rottgering03}), WSRT (Westerbork Synthesis Radio Telescope
\cite{oosterloo10}) and ASKAP (Australian SKA
Pathfinder \cite{Johnston08}) telescopes, and the prospect of the Square Kilometre Array (SKA) in the coming decade. Increased sensitivity, a very wide sky coverage,
and deep redshift reach will facilitate cosmological observations
with significant accuracy.

This has only recently been explored in
\cite{Raccanelli:2011pu}, which analyzed what can be achieved by surveys with LOFAR \citep{rottgering10}, WSRT (WODAN, Westerbork Observations of the Deep Apertif Northern
sky survey \citep{rottgering11})
and ASKAP (EMU, Evolutionary Map of the Universe \cite{Norris11}), via three experiments:
auto-correlation of radio sources, cross-correlation of radio
sources with the Cosmic Microwave Background (the late Integrated
Sachs-Wolfe effect), and cross-correlation of radio sources with
foreground objects (cosmic magnification). The auto-correlation function has been further investigated by \cite{camera}, which examines the impact of cross-identification of radio sources with optical redshift surveys.

The huge volumes covered by forthcoming radio surveys, and their deep redshift reach in comparison to current and future optical surveys, mean that correlations on scales above the Hubble horizon $H^{-1}(z)$ will be measured. On these scales, the standard analysis of the power spectrum is inadequate -- because this analysis is Newtonian, i.e. it is based on the assumption of sub-Hubble scales. The Newtonian analysis must be replaced by the correct general relativistic (GR) analysis in order to consistently incorporate super-Hubble scales. On small scales, the Newtonian analysis is a very good approximation, but on larger and larger scales, the GR corrections become more significant.

On these larger scales, any primordial non-Gaussianity in the matter distribution also grows larger. This probe of non-Gaussianity is expected to become competitive with the CMB for large-volume surveys such as those in the radio. Thus it is important to perform a GR analysis in order to correctly identify the non-Gaussian signal.

Unfortunately cosmic variance also becomes more and more of a problem on these larger scales covered by radio surveys. However, it is possible to beat down cosmic variance by using different tracers of the underlying matter distribution.

In this paper, we re-visit the Newtonian analysis of radio continuum surveys, and include for the first time the terms that were not considered in the auto-correlations computed by \cite{Raccanelli:2011pu,camera} -- i.e. redshift space distortions, lensing convergence and the GR corrections (potential and velocity terms). We do this first in the Gaussian case, and then when there is primordial non-Gaussianity of the local type.

\section{General Relativistic corrections to the angular power
spectrum}

A GR analysis of the matter power spectrum needs to start by
correctly identifying the galaxy overdensity $\Delta$ that is {\em
observed} on the past light cone. In the standard Newtonian
approach, the overdensity $\delta$ is defined in some gauge. A
change of gauge gives effectively the same results on sub-Hubble scales, but
leads to different results on large scales -- and this remains
true even if we use gauge-invariant definitions of $\delta$. There are many gauge-invariant definitions of the overdensity, but the
observed  $\Delta$ is unique, and is necessarily gauge-invariant.
In addition, we need to account for the distortions arising
from observing on the past light cone, including all redshift space effects and volume distortions.

\begin{figure*}[htp]
\begin{center}
\epsfig{file=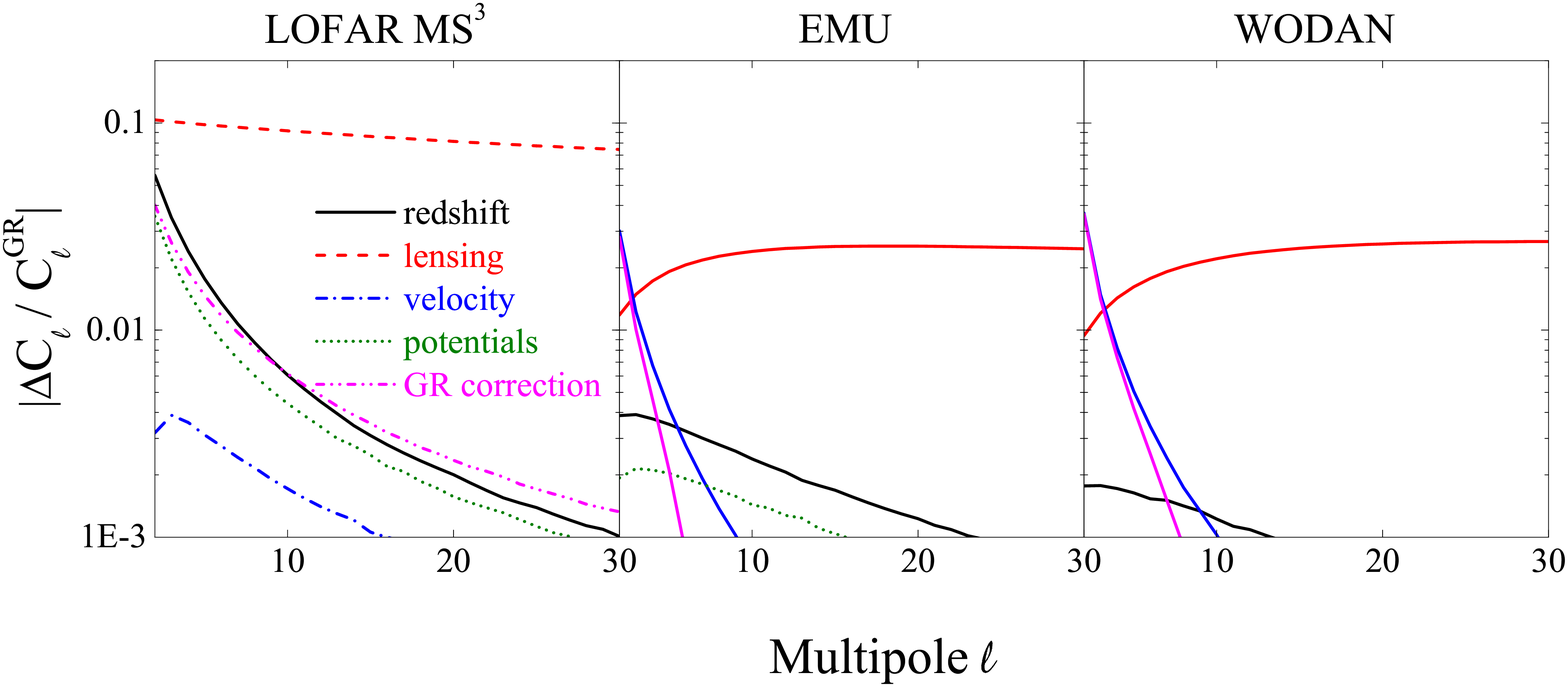, width=1.15\linewidth}
\caption{Fractional error $\Delta C^X_{\ell}/C^{\rm GR}_{\ell}$ defined in (\ref{delcl}), when the various types of terms X in the full result $ C^{\rm GR}_{\ell}$ are neglected. This is shown for the LOFAR MS$^3$, EMU and WODAN continuum surveys. Here and in Fig. \ref{cv}, a broken curve denotes the absolute value of a negative quantity. The GR correction for LOFAR is shown on its own in Fig.~\ref{cv}.} \label{lew}
\end{center}
\end{figure*}

The GR analysis of the power spectrum has recently been developed~\cite{Yoo:2010jd,Yoo:2010ni,Bonvin:2011bg,Challinor:2011bk,Bruni:2011ta,
Baldauf:2011bh,Jeong:2011as,Yoo:2011zc,LopezHonorez:2011cy,schmidtrulers,jeonggrav,schmidtgravshear,Yoo:2012se}, and the consequences for the correlation function at wide angles and large scales have also been computed~\cite{bertaccaGR}.
The observed overdensity in direction ${\bm{n}}$ at redshift $z$ is
 \ba
\Delta({\bm{n}},z) =\delta_z({\bm{n}},z) +{\delta V({\bm{n}},z) \over V(z)},
 \ea
where $\delta_z$ is the redshift space density perturbation and $V$ is the physical volume density per redshift interval per solid angle~\cite{Bonvin:2011bg}. Each term on the right is physically defined and hence gauge-invariant. In order to compute these terms, we can choose any gauge that we prefer. Here we work in the Newtonian gauge,
\begin{equation}
ds^2= a(\eta)^2 \Big[- (1+2 \Phi) d \eta^2 + (1-2\Psi) d \bm{x}^2
\Big].
\end{equation}

\begin{figure}
\begin{center}
\epsfig{file=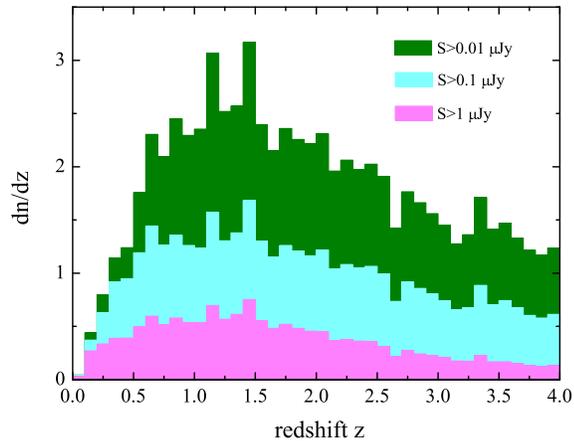, width=\linewidth}
 \caption{Number
density for SKA-like continuum survey with different sensitivities $S$, in units (arcmin)$^{-2}$. (The number counts $W$ in \eqref{alm} are per solid angle.)
} \label{skanz}
\end{center}
\end{figure}

In the Newtonian approach, the standard observed overdensity includes redshift space distortion and lensing contributions~\cite{Matsubara}. Lensing is often omitted in the Newtonian approach, although the lensing contribution to the density contrast can be significant \cite{Namikawa:2011yr}. 
In their analysis of radio continuum surveys, \cite{Raccanelli:2011pu} did not include either the redshift space distortions or lensing  -- and our angular power spectra here generalize their results also within the standard Newtonian approximation. The standard observed overdensity is
 \ba
\Delta^{\rm std}= \delta - \frac{1}{\H} {\bm{n}}
\cdot \frac{\partial \bm{v}}{\partial \chi}  +(5s-2) \kappa,  \label{deltastd}
 \ea
where $\chi$ is the comoving radial position of the source,
$\bm{v}$ is the peculiar velocity, and $\kappa$ is the lensing
convergence,
 \be
\kappa=-{1\over2}\nabla_{{\bm{n}}}^2\int_{\eta_{\rm o}}^\eta
d\tilde\eta \, {(\tilde \eta-\eta) \over (\eta_{\rm
o}-\eta)(\eta_{\rm o}- \tilde\eta)}(\Phi+\Psi) .
 \ee
The slope $s$ characterizes the change of the number density with respect to the threshold magnitude $m_{\ast}$ \cite{Challinor:2011bk}:
 \be\label{eq:slope} 
 s\equiv \f{\partial}{\partial m_{\ast}} {\rm log_{10}}N(z,m<m_{\ast})= {2 \over 5}\big(\langle \alpha-1 \rangle+1\big),
 \ee
where the parameter $\langle \alpha-1 \rangle$ is used in \cite{Raccanelli:2011pu}.

The observed overdensity in GR is given by~\cite{Challinor:2011bk,Bonvin:2011bg}
 \ba
&& \Delta^{\rm GR} = \Delta^{\rm std}+ \Delta^{\rm GR\, corr},
\label{full} \\
&& \Delta^{\rm GR\, corr}= (A+1)\Phi +(5s-2)\Psi +
\frac{1}{\H}{\Psi}' \nonumber\\
&&~~ + A\int^{\eta_{\rm o}}\!d \eta ({\Phi}' + {\Psi}') 
 + \frac{(2-5s)}{\chi} \int^{\eta_{\rm o}}\! d\eta(\Phi +
\Psi) \nonumber\\
&&~~ -A\, {\bm{n}}\cdot
\bm{v} , \label {deltan}
 \ea
where
 \be
A \equiv {\H' \over \H^2}+\frac{2-5s}{\H\chi} +5s.
 \ee
The GR correction includes potential contributions (both local at the source and integrated along the line of sight) and a Doppler velocity contribution.

A further gauge subtlety arises in relation to the bias of the
sources. The simple bias relation $\delta=b \delta_m$, where $b$ depends only on redshift and not on scale, may be
applied in any gauge on sub-Hubble scales. But on larger scales,
the relation is gauge-dependent, and we need a GR analysis. The
simple bias law is applicable in the synchronous-comoving gauge,
as shown by analysis of the spherical collapse model and by the
physical argument that galaxies and dark matter follow the
same velocity field
\cite{Bonvin:2011bg,Challinor:2011bk,Bruni:2011ta,Baldauf:2011bh}.
Then we must transform from synchronous-comoving  in order to get the correct bias relation in Newtonian gauge:
\begin{equation}
 \delta = b \delta_m + 3\frac{a H}{k } (b-1) {v} \,.
 \label{gal-Newton}
\end{equation}

\section{Observable effects of GR corrections}

\begin{figure*}
\begin{center}
\epsfig{file=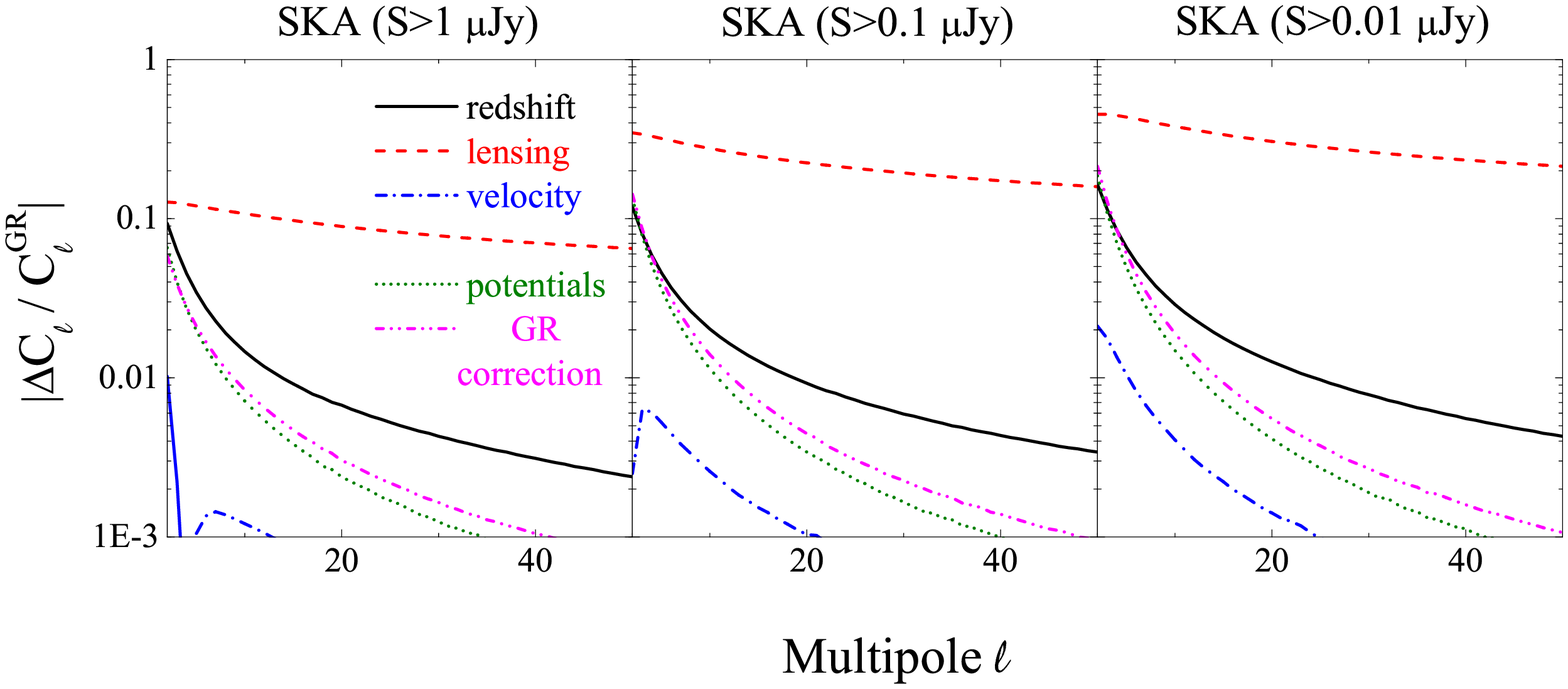, width=1.15\linewidth}\caption{
The same as Fig.~\ref{lew}, for future SKA-like continuum surveys with different flux limits $S$. For the highest
sensitivity, the GR correction is shown on its own in Fig.~\ref{cv}.} \label{skas}
\end{center}
\end{figure*}
\begin{figure*}
\begin{center}
\epsfig{file=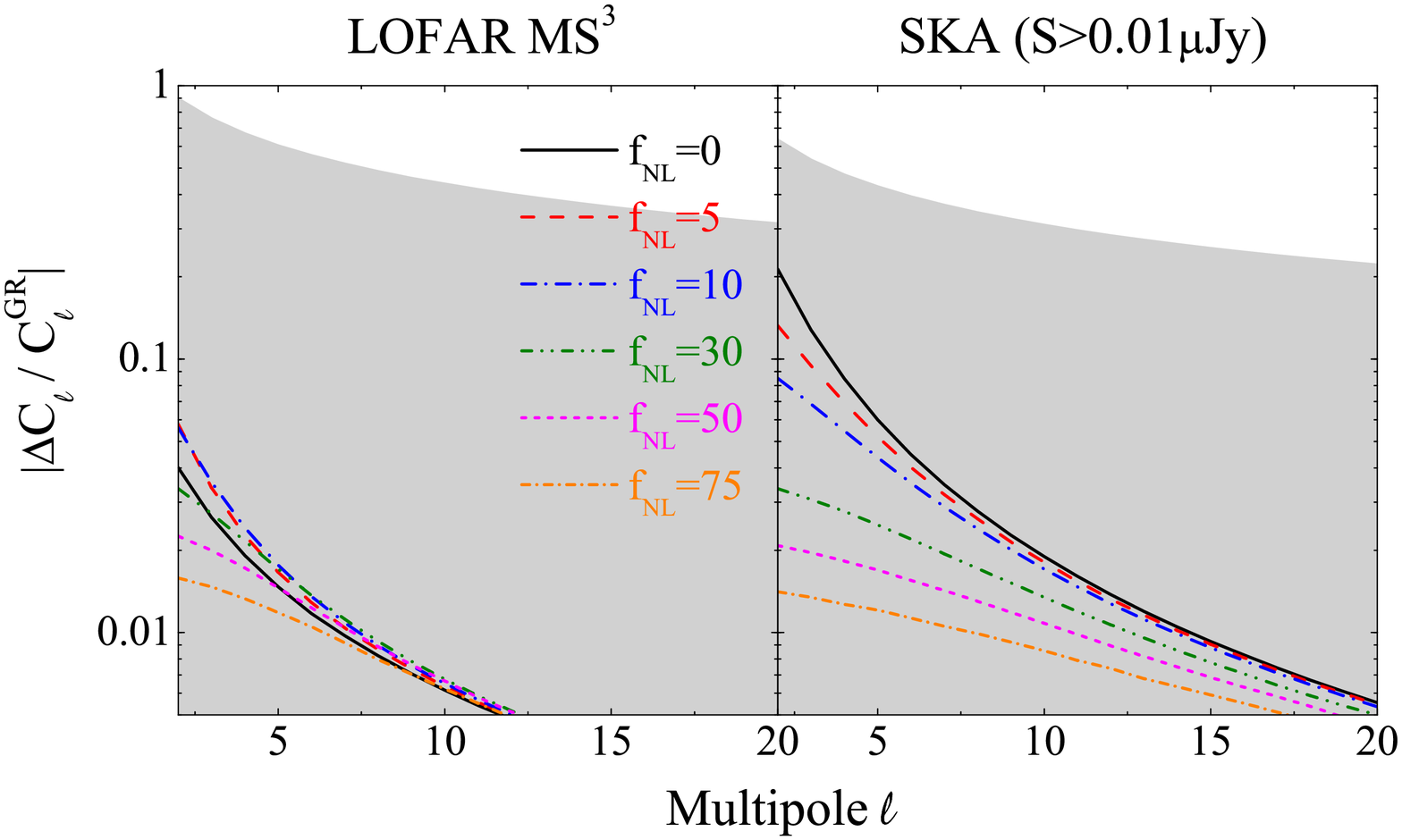, width=1.\linewidth}\caption{
The fractional GR corrections,  $|\Delta C^{\rm GR\, corr}_\ell(f_{\rm NL})/C^{\rm GR}_\ell(f_{\rm NL})|$, for LOFAR MS$^3$ and ultra-SKA ($S>0.01\mu$Jy) continuum surveys, including cases with non-Gaussianity. Cosmic variance is shown by shaded regions. (Note that the correction is negative in all cases.)} \label{cv}
\end{center}
\end{figure*}

In a radio continuum survey, sources are detected via a standard approach, as used in \cite{Raccanelli:2011pu,camera}.
First the imaging is done (using e.g. CLEAN, maximum entropy, or sparsity methods), and then discrete objects are detected using SExtractor or similar codes.
(This is to be distinguished from intensity mapping.) The sources that are isolated from images generated from the survey are used to compute the power spectrum.
Since the surveys integrate over redshift, the overdensity in a given direction may be expanded in
spherical harmonics as
 \ba
\Delta( \bm{n}) &=& \sum_{\ell m} a_{\ell m} Y_{\ell m}(\bm{
n}),\\a_{\ell m}&=&\int d\Omega_{\bm{n}}\,dz\, Y^*_{\ell m}(\bm{n}) W(z) \Delta(\bm{n},z),
\label{alm}
 \ea
where $W(z)$ is the number count per unit solid angle at redshift $z$.
Then the integrated angular power spectrum is
 \ba
C_\ell  &=&  \langle |a_{\ell m}|^2\rangle \label{iaps} \\
&=& {2\over \pi}\!\int \! d\ln k\, dz\, dz' \, {\cal P}_{\rm in}(k)
F_\ell(k,z)W(z)F^*_\ell(k,z')W(z'), \nonumber
 \ea
where ${\cal P}_{\rm in}$ is the (dimensionless) initial power
spectrum of $\Phi$, and the function $F_\ell$ is of the form \cite{Bonvin:2011bg}
 \ba
&& F_\ell(k,z)= j_\ell(k\chi)f_1(k,z)+ j'_\ell(k\chi)f_2(k,z)\nonumber\\
&&~~~~~+j''_\ell(k\chi)f_3(k,z)+ \int d\chi \, j_\ell(k\chi)f_4(k,z).
 \ea
Here the $f_a$ incorporate the auto- and cross-correlations of the various terms in \eqref{deltastd} and \eqref{deltan}.

We can compute \eqref{iaps} with and without GR corrections in order to assess their impact. It is
useful to define (following \cite{Bonvin:2011bg,Challinor:2011bk})
 \be \label{delcl}
\Delta C_\ell^{\rm X} = C_\ell^{\rm GR}- C_\ell^{\rm GR - X},
 \ee
where GR denotes the $C_\ell$ calculated with the full GR overdensity \eqref{full} and X denotes the contribution from various terms in
\eqref{deltastd} and (\ref{deltan}):
\ba
X= && \mbox{redshift term } (\propto \partial \bm{n}\cdot\bm{v}/\partial\chi),\nonumber\\ && \mbox{lensing term } (\propto \kappa),\nonumber\\&& \mbox{velocity term } (\propto \bm{n}\cdot \bm{v}),\nonumber\\&& \mbox{potential terms (all those involving } \Phi, \Psi),\nonumber\\&& \mbox{GR correction term } \Delta^{\rm GR\, corr}. \nonumber
\ea
Therefore $\Delta C^X_\ell$ gives
the contribution of X in each case.
A negative $\Delta C_\ell^{\rm GR\, corr}$ means that GR corrections
reduce $C_\ell$.

\begin{figure*}
\begin{center}
\epsfig{file=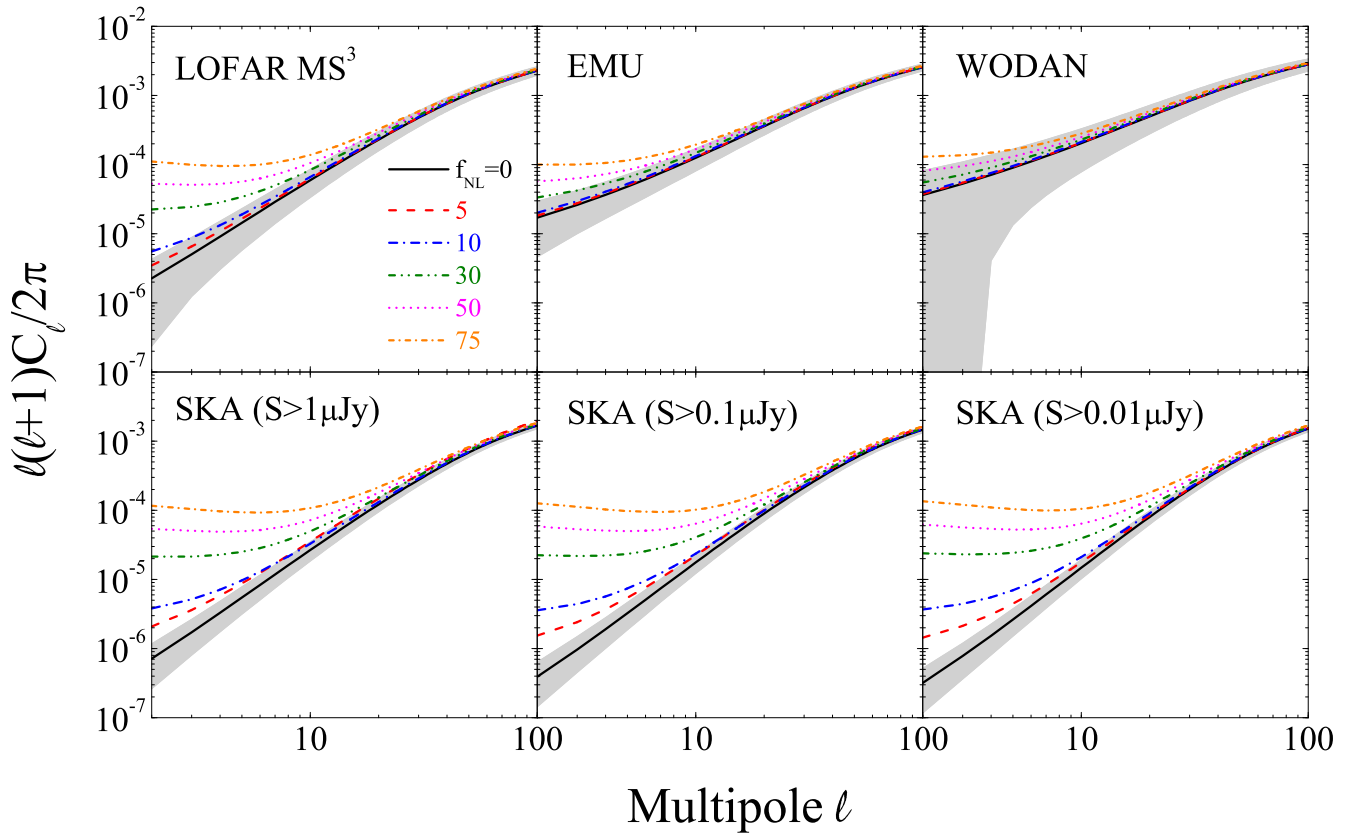, width=1.1\linewidth}
\caption{The imprint of primordial non-Gaussianity on the observed galaxy-galaxy angular auto-correlation spectra for various radio continuum surveys, including full GR corrections. Shaded bands show the cosmic variance. } \label{nonG}
\end{center}
\end{figure*}
\begin{figure*}
\begin{center}
\epsfig{file=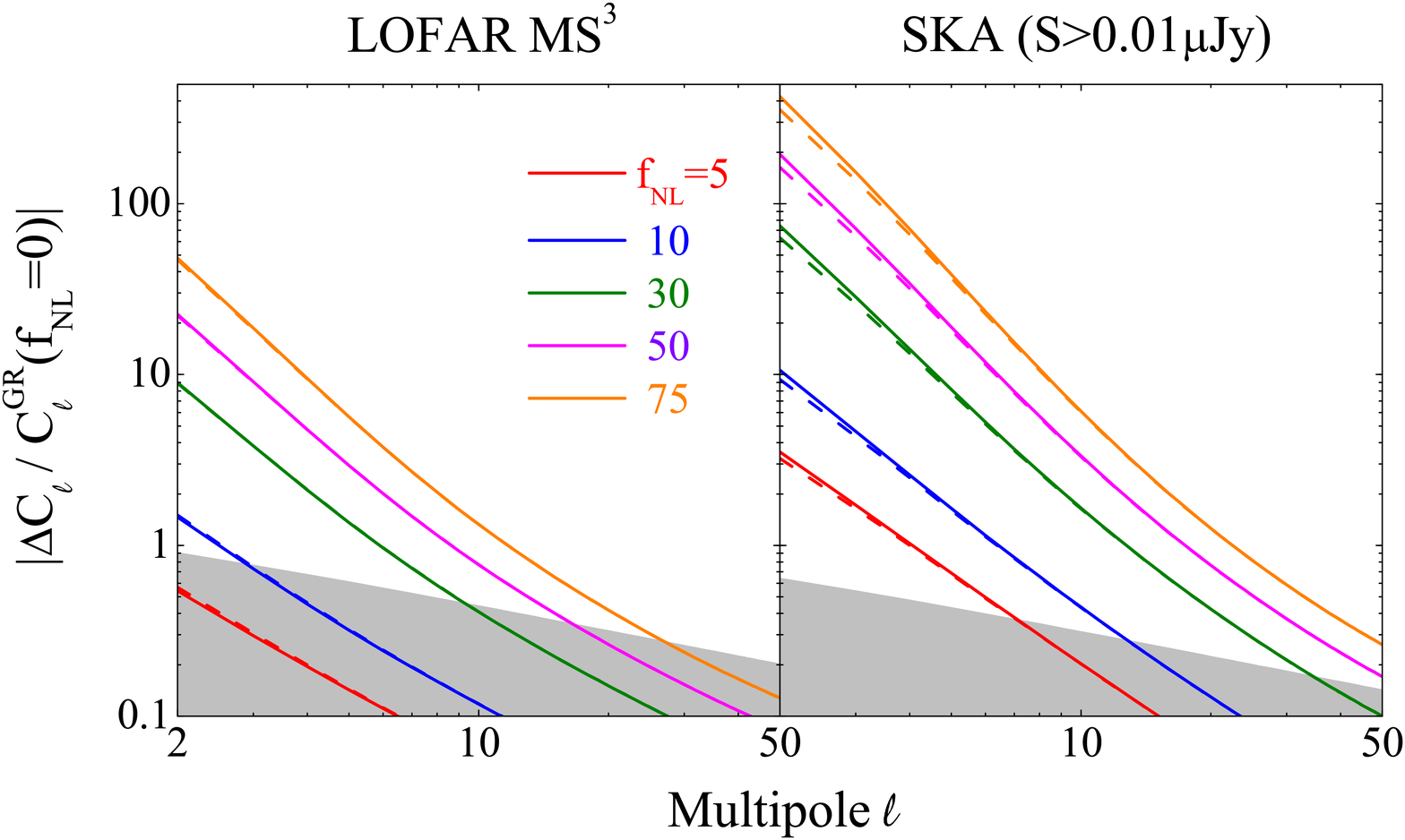, width=\linewidth}\caption{
The fractional difference of the galaxy count spectra between the cases with and without non-Gaussianity, $ |C_\ell(f_{\rm NL})-C_\ell(f_{\rm NL}=0)|/ C_\ell^{\rm GR}( f_{\rm NL}=0)$. The solid and dashed curves illustrate the spectra with and without GR corrections respectively. Shaded regions show the cosmic variance.} \label{nonG_ratio}
\end{center}
\end{figure*}

\subsection*{LOFAR, EMU, WODAN}

We adopt the models of \cite{Raccanelli:2011pu} for radio source number density $n(z)$ and bias $b(z)$ for the LOFAR, EMU and WODAN surveys and
we use their luminosity function parameter $\langle \alpha - 1 \rangle (=2.5(s-1))$ for each survey. (See Fig. 1 in \cite{Raccanelli:2011pu} for $n(z)$.)
For illustrative purposes, we use the higher flux density threshold LOFAR MS$^3$ number densities; the LOFAR Tier 1 results are similar to the EMU results. 
Assuming Gaussian perturbations, for the surveys studied in
\cite{Raccanelli:2011pu}, the various terms in \eqref{deltastd} and (\ref{deltan}) are shown in Fig.~\ref{lew}. The curves are the absolute value of $\Delta C_{\ell}^{\rm X}$ defined in (\ref{delcl}), rescaled by the spectra with full GR corrections $C_{\ell}^{\rm GR}$. 

The GR corrections are percent-level at low multipoles $\ell \lesssim \,$few, and are largest in the LOFAR MS$^3$ case.

\subsection*{SKA}

We obtain models of $n(z)$ for SKA-like continuum surveys (i.e. without redshifts) from the $S^3$ simulations of~\cite{wilman}. We examine both moderate and ultra-deep SKA-like surveys, with flux density thresholds of 1$\mu$Jy, 0.1 $\mu$Jy and 0.01$\mu$Jy, applying the bias models of \cite{Raccanelli:2011pu} for each source type. This highest sensitivity survey will be costly to achieve, but we examine it to see whether GR corrections are important in the extreme case. In addition, we measure the luminosity function parameter $\langle \alpha - 1 \rangle (=2.5(s-1))$ using the $S^3$ simulations. We find $\langle \alpha-1 \rangle$ of $-0.20, -0.40$ and $-0.55$ (corresponding to a slope $s=0.32,0.24$ and $0.18$) for the $1\mu$Jy,  $0.1\mu$Jy and $0.01\mu$Jy surveys respectively.
The number density for various sensitivities is shown
in Fig.~\ref{skanz}. 

The resulting size of the various contributions and of the GR corrections is
shown in Fig.~\ref{skas}. For the highest sensitivity SKA, the GR
effects are at the $O(10)\%$ level for $\ell \lesssim \,$few. 

\subsection*{Cosmic variance}

Figure~\ref{cv} shows the GR correction signal within the cosmic variance (shaded bands):  for the LOFAR MS$^3$ (solid black in the left panel) and
for the highest sensitivity ultra-SKA ($S>0.01 \mu{\rm Jy}$) case where the GR correction is most visible (solid black in the right panel). The cosmic variance for a survey covering a fraction $f_{\rm sky}$ of the sky is
 \be
\sigma(C_{\ell})=\sqrt{\f{2}{(2\ell+1)f_{\rm sky}}}\,C_{\ell}.
 \ee
This is highest on the largest scales, where the GR corrections are strongest. 
It is not surprising that cosmic variance overwhelms the GR correction signal, as shown in Fig. \ref{cv}. However, cosmic
variance can be overcome by using multiple tracers of the underlying matter distribution \cite{Seljak:2008xr,McDonald:2008sh,White:2008jy}.
In principle the GR corrections in radio continuum surveys may be detectable using this method. Furthermore, if the observations include source redshifts -- i.e. if we have an HI survey -- then we expect that the GR corrections will be stronger than in the continuum case, where redshift is integrated over.

\section{Non-Gaussianity and GR corrections}

Several theories of the very early universe predict non-Gaussian initial conditions for the probability distribution function of cosmological perturbations. One way to test this is via the large-scale clustering of matter.
Among the numerous models of non-Gaussianity considered (see \cite{komatsu10} and references therein), one of the most studied is the so-called local form, in which deviations from Gaussianity are parametrized as:
\begin{equation}
\Phi = \phi_g - f_{\rm NL} \big(\phi_g^2 - \langle \phi_g^2
\rangle \big),
\end{equation}
where $\phi_g$ is Gaussian.
This leads to a scale-dependent correction
to the bias \cite{matarrese2000, dalal08, desjacques10, Xia:2010yu,Xia:2010pe,Bruni:2011ta}:
 \be
b(k,z)={b}_g(z)+f_{\rm NL}\big[{b}_g(z)-1\big] {3\delta_{\rm ec} \Omega_m H_0^2 \over
c^2k^2 T(k) D(z)}, \label{non-G-bias}
 \ee
where $b_g$ is the Gaussian bias, $\delta_{\rm ec}$ is the critical
matter overdensity for ellipsoidal collapse, $T(k)$ is the transfer function, $D(z)$ is the growing mode of density
perturbations, and $c$ is the speed of light. Following \cite{Xia:2010yu}, we take $\delta_{\rm ec}=1.68\sqrt{0.75}=1.45$.

In a radio continuum survey we count the number of galaxies (as opposed to intensity mapping), and so we expect that galaxy number density peaks are related to the high density peaks in the overall matter distribution in the usual way. Therefore it should be reasonable to apply \eqref{non-G-bias}.

A scale-independent bias with $f_{\rm NL}=0$, as in \eqref{gal-Newton},  becomes scale-dependent when  $f_{\rm NL}\neq 0$, by \eqref{non-G-bias}. Therefore the non-Gaussianity can in principle change the shape of the angular power spectrum on large scales. The non-Gaussian signal clearly grows with scale -- as does the GR correction. In order to obtain the correct prediction of non-Gaussianity, it is therefore necessary to use the full GR power spectrum, as pointed out in \cite{Bruni:2011ta}. Although the GR corrections are independent of non-Gaussianity, the full $C^{\rm GR}_\ell$ includes cross-correlations of GR correction terms with non-Gaussian correction terms, so that the difference between $C^{\rm GR}_\ell$ and $C^{\rm std}_\ell$ can vary with $f_{\rm NL}$.

Figure~\ref{nonG} shows the predicted $C_{\ell}$ with various values of $f_{\rm NL}$ for different radio surveys. We have included full GR corrections in all spectra.
We can see a clear deviation on large scales that overcomes cosmic variance for a non-Gaussianity signal of $f_{\rm NL}\gtrsim20$ in the LOFAR MS$^3$ case, and $f_{\rm NL}\gtrsim5$ for a SKA continuum survey at $1\mu$Jy or fainter.

This can also be seen in Fig.~\ref{nonG_ratio}, where the fractional differences between the cases with and without non-Gaussianity are plotted. Comparing the spectra with (solid) and without (dashed) GR corrections, it is clear that ignoring the GR correction has a marginal effect on the detectable non-Gaussian signal for the sensitivity of LOFAR MS$^3$,  and a small effect (though growing with increasing $f_{\rm NL}$ and scale) on ultra-SKA ($S>0.01\mu$Jy).

The presence of dominant non-Gaussianity has a sizeable effect on the detectability of GR corrections. As Fig.~\ref{cv} shows, in most cases a large $f_{\rm NL}$ makes the GR corrections relatively less important. This is understandable since $f_{\rm NL}$ only boosts the `standard' term (\ref{deltastd}) (via the overdensity term $\delta$), which makes the GR correction terms less visible.

It is however worth pointing out that in radio surveys the radial galaxy distribution and the bias are particularly uncertain. They are usually modeled starting from theoretical (e.g.~\cite{dp90,matarrese97}) or observational (e.g.~\cite{raccanelli08}) arguments.
In particular, the same effect of  increased power of angular correlations  on large scales can be due to non-Gaussianity~\citep{Xia:2010pe} or to a different model for the bias~\citep{raccanelli08}.
In order to make detailed predictions of the amount of non-Gaussianity that can be detected in forthcoming radio surveys, it will be necessary to model the effects of variations of $n(z)$ and $b(z)$, but this is beyond the scope of this paper and is left to a future work.

\section{Conclusions}

We have presented an analysis of GR corrections to the angular power spectrum of radio sources that will be measured with forthcoming radio continuum surveys such as LOFAR, EMU and WODAN, along with predictions for SKA-like surveys.
These surveys will be well suited for probing GR corrections to the standard Newtonian analysis of the spectra, since they will observe super-Hubble scales not yet surveyed, being very deep and wide.

We have computed for the first time the contributions from redshift distortions and lensing convergence to the angular power spectrum for radio continuum surveys, thus generalizing previous results that incorporated only the overdensity contribution \cite{Raccanelli:2011pu}. Then we have included the further GR corrections, in the form of velocity and potential terms. We have shown how all these contributions will, individually and in combination, affect measurements of the angular power spectrum for the different surveys considered.
The GR corrections to the standard Newtonian analysis are most significant on the largest scales, reaching $O(10\%)$ for the SKA. However, precisely because they grow with scale, they are dominated by cosmic variance for the near future SKA Pathfinder generation surveys, while they could be in principle observable by a sensitive enough SKA-like survey. With complementary information from other tracers (e.g. via the Euclid optical/ IR survey), cosmic variance can be overcome.

A large-scale increase in power can also be due to primordial non-Gaussianity. This means that a GR analysis is essential for a correct calculation of the non-Gaussian signal. We have computed the corrections arising from the local form of non-Gaussianity for the different surveys. Comparing the non-Gaussian effect to that of  GR corrections without non-Gaussianity, we find that non-Gaussian corrections to the power spectrum will dominate over GR corrections for continuum surveys. The non-Gaussian signal rises above cosmic variance on large enough scales as follows: for WODAN when $f_{\rm NL}\gtrsim80$; for EMU when $f_{\rm NL}\gtrsim50$; for LOFAR when $f_{\rm NL}\gtrsim20$ and for SKA when $f_{\rm NL}\gtrsim5$.
    
Continuum radio surveys do not provide redshift information, so that GR corrections can be degenerate with a change in the distribution of matter, given by the product of radial source distributions with the bias. We expect that an SKA HI galaxy survey will show stronger and more clearly defined GR corrections, because spectroscopic information will break this degeneracy.

\vspace{0.5in}

\noindent {\bf Acknowledgments:}\\
We thank Matt Jarvis for helpful discussions.
RM was supported by the South African Square Kilometre
Array Project and National Research Foundation. RM, GZ, DB, KK were supported by the UK Science \& Technology Facilities Council (grant no. ST/H002774/1) and by a Royal
Society (UK)/ National Research Foundation (SA) exchange grant. KK was also supported by the European Research Council and the Leverhulme Trust. Part of the research
described in this paper was carried out at the Jet Propulsion
Laboratory, California Institute of Technology, under a contract
with the National Aeronautics and Space Administration.

\end{document}